\documentclass{book}
\usepackage{amsfonts} 
\usepackage{graphicx} 

\newcommand{\la}{\langle}
\newcommand{\ra}{\rangle}

\newcommand{\beq}{\begin{eqnarray}}
\newcommand{\eeq}{\end{eqnarray}}

\newcommand{\mn}{{\mu\nu}}

\newcommand{\Dsl}{/ {\hskip-0.26cm{D}}}

\newcommand{\bp}{\mbox{\boldmath $p$}}

\newcommand{\br}{\mbox{\boldmath $r$}}
\newcommand{\brs}{\mbox{\scriptsize \boldmath $r$}}

\newcommand{\bks}{\mbox{\scriptsize \boldmath $k$}}

\newcommand{\qL}{q_{_{\rm L}} }
\newcommand{\qR}{q_{_{\rm R}} }
\newcommand{\bqL}{\bar{q}_{_{\rm L}} }
\newcommand{\bqR}{\bar{q}_{_{\rm R}} }
\newcommand{\qLR}{q_{_{\rm L(R)}} }

\newcommand{\dL}{d_{_{\rm L}} }
\newcommand{\dR}{d_{_{\rm R}} }
\newcommand{\dLR}{d_{_{\rm L(R)}} }

\newcommand{\als}{\alpha_{\rm s}}

\newcommand{\Lamqcd}{ \Lambda_{_{\rm QCD}} }

\def\simge{\mathrel{%
       \rlap{\raise 0.511ex \hbox{$>$}}{\lower 0.511ex \hbox{$\sim$}}}}
\def\simle{\mathrel{
       \rlap{\raise 0.511ex \hbox{$<$}}{\lower 0.511ex \hbox{$\sim$}}}}
\def\abf{a_{\rm bf}^{\ }}

\def\gbf{g_{\rm bf}^{\ }}

\def\aN{a_{_{\rm NN}}}

\def\mN{m_{_{\rm N}}}
\def\mb{m_{\rm b}^{\ } }
\def\mf{m_{\rm f}^{\ } }
\def\mR{m_{_{\rm R}} }

\def\mub{\mu_{\rm b}^{\ } }
\def\muf{\mu_{\rm f}^{\ } }

\def\bfm#1{\mbox{\boldmath $#1$}}
\newcommand{\BQ}{\begin{equation}}
\newcommand{\EQ}{\end{equation}}
\newcommand{\BQA}{\begin{eqnarray}}
\newcommand{\EQA}{\end{eqnarray}}

\title{\huge {\bf Chapter 25:\\  \   \\
Quantum Phase Transitions \\ in Dense QCD}\footnote{To appear in {\em Developments in
Quantum Phase Transitions}, ed. L. D. Carr  (Taylor and Francis, 2010).}}
\author{\rm T. Hatsuda and K. Maeda \\
Physics Department, The University of Tokyo,\\
 Tokyo 113-0033, Japan}

\begin{document}
\maketitle
\frontmatter

\pagestyle{myheadings}
\markboth{ }{Table of Contents}
\tableofcontents
\mainmatter
\markboth{ }{Quantum Phase Transitions in Dense QCD}

\setcounter{chapter}{24}
\chapter{Quantum Phase Transitions in Dense QCD}

    Quantum chromodynamics (QCD) at finite temperature, $T$, and quark chemical potential,
$\mu$, has a rich phase structure:  at low $T$ and low $\mu$, the Nambu-Goldstone
(NG) phase with nearly massless pions is realized by the dynamical breaking of
chiral symmetry through condensation of quark$-$anti-quark pairs,
  while, at low $T$ and high $\mu$, a Fermi liquid
of deconfined quarks is expected to appear as a consequence of asymptotic
freedom.  Furthermore, in such a cold quark matter, condensation of
quark$-$quark pairs leads to the color superconductivity (CSC).  At high $T$ for 
arbitrary
$\mu$, all the condensates melt away and a quark-gluon plasma (QGP)
is realized.  The experimental exploration of thermal phase transition
 from the NG phase to QGP
 is being actively pursued in ultrarelativistic heavy ion
collisions at RHIC (Relativistic Heavy Ion Collider),
 and will be continued in the future at LHC (Large Hadron Collider).
   The quantum phase transition from the NG phase to the CSC at low $T$ 
   is also relevant to heavy-ion
collisions at moderate energies, and is of
interest in the interiors of neutron stars and possible quark stars.

In this Chapter, after a brief introduction to the basic properties of QCD,
  the current status of the QCD phase structure and associated 
  quantum phase transitions  will be summarized with particular
   emphasis on the symmetry realization of each phase.  
   Possible connection between the physics of QCD and that of ultracold atoms 
 is also discussed.

\section{Introduction to QCD}
\label{sec:intro}

The color SU(3)$_{\rm C}$ gauge theory
 of quarks and gluons \cite{Nambu:1966} is now called
  the quantum chromodynamics (QCD) and is
 established as  the fundamental theory of strong interaction.
 The Lagrangian density of QCD reads
\beq
\label{eq:QCDaction-LR}
{\cal L}_{\rm QCD} =   
   \bqL i\  \Dsl \qL + \bqR i\  \Dsl \qR \
    - \frac{1}{4} G_{\mn}^{\alpha} G^{\mn}_{\alpha} 
       +   \bqL m \qR + \bqR m \qL  ,
\eeq 
 where the covariant derivative is defined as 
  $D^{\mu} \equiv \partial^{\mu} + 
  i g t^{\alpha} {\cal A}^{\mu}_{\alpha}$ with 
  $g$ being the QCD coupling constant, $t^{\alpha}$ the 
  SU(3)$_{\rm C}$ group generator and  ${\cal A}^{\mu}_{\alpha}$ the 
 gluon field belonging to the adjoint representation of 
   SU(3)$_{\rm C}$. 
   The gluon field-strength tensor is
  defined as $G^{\mn}_{\alpha} \equiv \partial^{\mu}{\cal A}^{\nu}_{\alpha} -
 \partial^{\nu}{\cal A}^{\mu}_{\alpha}
  - gf^{\alpha \beta \gamma} {\cal A}^{\mu}_{\beta} {\cal A}^{\nu}_{\gamma} $
   with $f^{\alpha \beta \gamma}$
  being the structure constant of the SU(3)$_{\rm C}$ group.

 The quark field $q$ belongs to the fundamental representation of 
   SU(3)$_{\rm C}$.
   The right (left) handed quark $\qR=\frac{1}{2}( 1+\gamma_5)q$
   ($\qL=\frac{1}{2}(1-\gamma_5)q$) is  an eigenstate of the
   chirality operator $\gamma_5$ with the eigenvalue $+1\ (-1)$. 
   Although quarks have six flavors (u,d,c,s,t,b) in the real world \cite{KM:1973},
 we focus only on three light quarks (u,d,s)  in this Chapter,
 so that the quark mass matrix is  $m= {\rm diag} (m_{\rm u},m_{\rm d},m_{\rm s})$.
  As is evident from Eq.(\ref{eq:QCDaction-LR}), only the mass term
 can mix the left-handed quark and the right-handed quark in the QCD
 Lagrangian.

\begin{figure}
\begin{center}
\includegraphics[width=0.5\columnwidth]{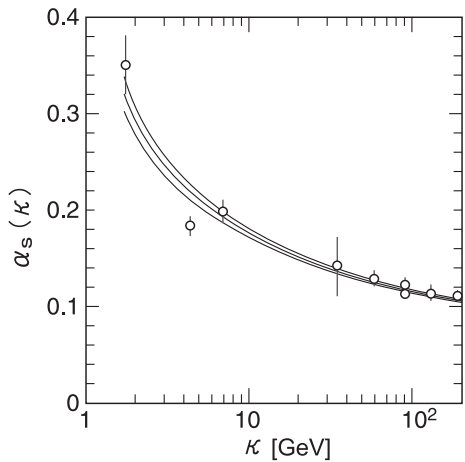}
\end{center}
\caption{The QCD fine-structure constant $\als$ determined  from 
 the $\tau$ decay,  the  $\Upsilon$ decay, the  deep inelastic scattering,
 the ${\rm e}^+{\rm e}^-$ 
 annihilation and the $Z$-boson resonance shape and width \cite{Amsler:2008zzb}. }
\label{fig:alpha_s} 
\end{figure}

 The running coupling constant
 $g(\kappa)$ is defined as an effective coupling strength
  at the energy scale $\kappa$.
 Due to the asymptotic freedom of
 QCD
 \cite{Gross:2005kv}, 
 $g(\kappa)$ becomes small when $\kappa$ increases as
 seen explicitly in the two-loop perturbation,
\beq
\label{eq:run-g}
\als  (\kappa) = \frac{g^2(\kappa)}{4\pi} \simeq
\frac{1}{4\pi \beta_0 \ln (\kappa^2 / \Lamqcd^2 ) }  \cdot
 \left[ 1 -  \frac{\beta_1}{\beta_0^2 } 
 \frac{\ln (\ln (\kappa^2/\Lamqcd^2))}
{\ln(\kappa^2/\Lamqcd^2)} \right] ,
\eeq 
where 
 $\beta_0= (11-\frac{2}{3}N_f)/(4\pi)^2$ and 
 $\beta_1=(102-\frac{38}{3}N_f)/(4\pi)^4$ with $N_f$ being the number of flavors.
 Here, $\Lamqcd (\simeq 200 \ {\rm MeV}=2 \times 10^8 \ {\rm eV})$
  is called the QCD scale parameter which is  determined from the comparison of 
  Eq.(\ref{eq:run-g}) with the experimental data in high energy
   processes satisfying $\kappa \gg \Lamqcd$ (see Fig.\ref{fig:alpha_s}).

 Equation (\ref{eq:run-g})
 implies that $\als (\kappa \sim \Lamqcd) \sim O(1) $, 
 so that the QCD  perturbation theory breaks down.
 This leads to various non-perturbative phenomena
 such as the confinement of quarks and gluons
  and the dynamical breaking of
  chiral symmetry \cite{Nambu:1961tp,Hatsuda:1994pi}
  at low energies, $\kappa < \Lamqcd $. These are responsible for the  
   formation of hadrons ($q\bar{q}$ mesons and 
    $qqq$ baryons), and also for the 
    origin of their masses.  
  On the other hand,   at extremely high temperature 
  and/or high baryon density where $\als(\kappa \gg \Lamqcd) \ll 1$, 
  the system may be treated as a 
   weakly interacting matter of quarks and gluons.
   Thus, there must be a phase transition 
    from the hadronic matter composed of confined quarks and gluons
     at low energies to the deconfined quark-gluon matter at high energies.

\begin{figure}
 \begin{center}
\includegraphics[width=0.85\columnwidth]{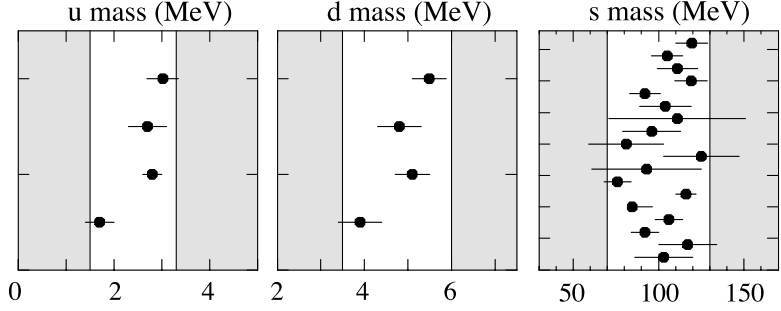}
 \end{center}
\caption{The masses of u, d and s quarks at the scale 
$\kappa=2\ {\rm GeV}= 2 \times 10^9 \ {\rm eV}$ determined from 
various observables and methods \cite{Amsler:2008zzb}. }
\label{fig:quark-mass_uds} 
\end{figure}

  Due to quantum corrections, the quark mass $m$ also
   becomes $\kappa$ dependent. 
  As seen from Fig.\ref{fig:quark-mass_uds}, the current determination of 
   the u and  d quark masses at $\kappa=2$ GeV indicates  that they are about 50 to 100 
   smaller than  $\Lamqcd$, while s quark mass
    is comparable to $\Lamqcd$.  
  Therefore, it is legitimate to treat $m_u/\Lamqcd$ and $m_d/\Lamqcd$
   as small expansion parameters, while the expansion by $m_s/\Lamqcd$    
  does not necessarily work.  Systematic expansion in terms of the 
   quark masses is called the chiral perturbation theory and 
    has been successfully applied to a wide variety of QCD phenomena \cite{Weinerg:2009}.

\subsection{Symmetries in QCD}
Let us consider the following transformations of the quark fields,
\beq
 \qL \rightarrow {\rm e}^{-i\theta_{\rm B}}\ {\rm e}^{-i\theta_{\rm A}}
  \ V_{\rm L}\ V_{\rm C}\  \qL, \  \  \ \ 
 \qR \rightarrow {\rm e}^{-i\theta_{\rm B}}\ {\rm e}^{+i\theta_{\rm A}}
 \ V_{\rm R}\  V_{\rm C}\  \qR,
\label{eq:quark-rotation} 
\eeq
where $V_{\rm C}$ (gauge rotation) is a local SU(3)$_{\rm C}$ transformation
 in the color space,
while $V_{\rm L(R)}$ (chiral rotation) 
is a global SU(3)$_{\rm L(R)}$ transformation in the flavor space.
The $\theta_{\rm B}$ and $\theta_{\rm A}$
 are phases associated with a global U(1)$_{\rm B}$ transformation (baryon-number rotation) 
 and the global U(1)$_{\rm A}$ transformation (axial rotation), respectively.
For $m_{\rm u,d,s}=0$ (the flavor-SU(3) chiral limit), QCD Lagrangian Eq.(\ref{eq:QCDaction-LR}) 
is invariant under Eq.(\ref{eq:quark-rotation}) together with the 
 SU(3)$_{\rm C}$ gauge transformation of the gluons, 
  so that 
 the full continuous symmetries of QCD become
\beq
\label{eq:GG-symmetry}
{\cal G} \equiv \bigl[ {\rm SU(3)}_{\rm C} \bigl]_{\rm local} 
\otimes \bigl[ {\rm SU(3)}_{\rm L} \otimes {\rm SU(3)}_{\rm R} \bigl]_{\rm global} 
\otimes \bigl[ {\rm U(1)}_{\rm B} \bigl]_{\rm global}.
\eeq
 Although the U(1)$_{\rm A}$  
 looks like a symmetry of 
  Eq.(\ref{eq:QCDaction-LR}), it is explicitly broken  by quantum effect
   known as the axial anomaly \cite{Treiman:1985} which
   reduces U(1)$_{\rm A}$  down to its discrete subgroup $Z(2N_f)_{\rm A}=Z(6)_{\rm A}$.  
 The masses of light quarks $m_{\rm u,d,s}$ act as small external fields to 
 break the global chiral symmetry
 $\bigl[ {\rm SU(3)}_{\rm L} \otimes {\rm SU(3)}_{\rm R} \bigl]_{\rm global}$. 

\begin{figure}[t]
\begin{center}
\includegraphics[width=0.7\columnwidth]{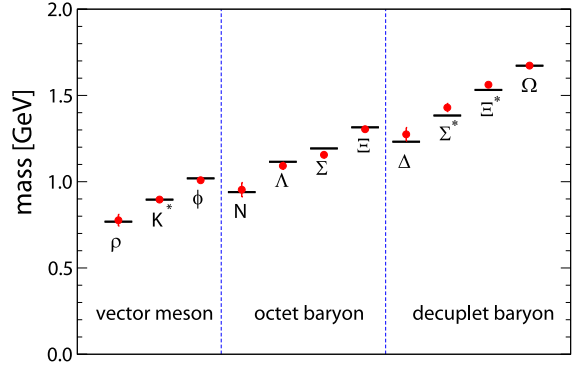}
\end{center}
\caption{Light hadron spectra obtained from lattice QCD Monte Carlo simulations
 with dynamical u, d, s quarks in the Wilson fermion formalism.
The spatial lattice volume $V$ and the lattice spacing $a$ are
 $(2.9\ {\rm fm})^3$ and $0.09\ {\rm fm}$, respectively.
Horizontal bars denote the experimental values \cite{Aoki:2008sm}. }
\label{fig:lattice-spect} 
\end{figure}

In the past few years, there arises a remarkable progress in 
lattice gauge theory \cite{Wilson:2004de} particularly  
 in calculating the hadron spectra on the basis of the QCD
  Monte Carlo simulations with light dynamical u, d, s quarks.
 This has been achieved partly due to the 
 growth of the supercomputer speed  and partly due to the
   new algorithms: Simulations with quark masses very close to the physical 
   values are now possible in the Wilson fermion formalism 
   \cite{Aoki:2008sm,Durr:2008xx}.
  Shown in Fig.\ref{fig:lattice-spect}  is an example of 
 such calculations for  meson and baryon masses
 extrapolated to the physical quark masses using the simulation data 
 taken in the interval,
  $\frac{1}{2}(m_{\rm u}+m_{\rm d}) =3.5\ {\rm MeV}-67\ {\rm MeV}$
  at $\kappa= 2~{\rm GeV}$. 
   The experimental values are reproduced in 3\% accuracy at present.

\subsection{Dynamical breaking of chiral symmetry}
\label{subsec:DBCS}

Although QCD Lagrangian in the flavor-SU(3) chiral limit has the symmetry ${\cal G}$
 in Eq.(\ref{eq:GG-symmetry}),
 the ground state of the system breaks some of the symmetries dynamically. 
 Consider the QCD vacuum $|0 \ra$ at zero temperature and zero
 baryon density.
   Taking into account the fact that the
   QCD does not allow dynamical breaking of parity and vector  symmetries 
   in the vacuum \cite{Vafa:1983tf},
  the following  is one of the  possible symmetry breaking patterns,
\beq
\label{eq:SB-pattern} 
{\cal G} 
  \rightarrow {\rm SU(3)}_{\rm {C}} \otimes 
  {\rm SU(3)}_{\rm {L+R}}  \otimes {\rm U(1)}_{\rm {B}}  ,
\eeq
where simultaneous
 transformation of the left-handed and right-handed quarks (the vector rotation, 
$V_{\rm L}=V_{\rm R}$) as indicated by ${\rm SU(3)}_{\rm {L+R}}$ 
remains as a symmetry of the vacuum. 
 An order parameter to characterize this ``dynamical breaking of chiral symmetry"  
 would be the ``chiral condensate", $\langle \bar{q}q \rangle_0 \equiv 
 \langle 0 | \bar{q}q |0 \rangle$, which is not invariant under
 the opposite
 rotation of the left-handed and right-handed quarks, 
 $V_{\rm L}=V_{\rm R}^{\dagger}$.
  Recent lattice QCD simulation  using
  overlap Dirac fermion
   with dynamical u,  d,  s quarks  indeed shows that Eq.(\ref{eq:SB-pattern}) 
 is realized with the chiral condensate \cite{:2009fh} 
\beq
\label{eq:quark-cond}
  \frac{1}{2}
\la \bar{\rm u}{\rm u}  + \bar{\rm d}{\rm d} \ra_0 \simeq 
- (242(04) \ {\rm MeV})^3 \ \ {\rm at}\  \kappa = 2 \ {\rm GeV}.  
\eeq
 The non-vanishing chiral condensate,
  $\la \bar{q}q \ra_0=\la \bqL \qR + \bqR \qL  \ra_0 \neq 0$,
  implies that the 
 quark$-$anti-quark pairs are  Bose-Einstein condensed. It also implies that
  a non-perturbative mixing between
   the left-handed and the right-handed quarks takes place in the QCD vacuum:
    In other words,  an effective quark mass called the chiral gap is dynamically generated.
  Indeed,  there are  phenomenological evidences
  that the u, d quarks and s quark have effective masses
   $M_{\rm u,d} \sim 350 $ MeV and $M_{\rm s} \sim 550$ MeV 
   inside hadrons \cite{Hatsuda:1994pi}.  

  The Nambu-Goldstone (NG) bosons associated with the dynamical breaking of
 the  flavor-SU(3) chiral  symmetry  are nothing but the pions, kaons and the $\eta$-meson.
  Moreover, one can derive a spectral sum rule, called the 
   Gell-Mann$-$Oakes$-$Renner relation \cite{GellMann:1968rz}, 
    which relates the pion mass to the chiral condensate as
$f_{\pi}^2 m_{\pi^{\pm}}^2 =
 - \hat{m} \la \bar{\rm u}{\rm u} + \bar{\rm d}{\rm d} \ra_0
 + O(\hat{m}^2) .$
 Here
  $\hat{m}\equiv(m_{\rm u}+m_{\rm d})/2$, 
  $f_{\pi}$( = 92.4 MeV) is the pion decay constant,
 and $m_{\pi^{\pm}}$ ($\simeq 140$ MeV) 
  are the charged  pion masses. Similar relation holds also for the 
   neutral pion $\pi^0$. In the limit $m_{\rm u,d} \rightarrow 0$, 
    the pion mass vanishes as it should be from the Nambu-Goldstone theorem.

\section{QCD matter at high temperature }

As the temperature $T$ of the system increases,
 the condensed $q\bar{q}$ pairs in the QCD vacuum
 are melted away by thermal fluctuations. This is 
    analogous to the phase transition in metallic superconductors
    with the electron pairing
     $\langle e_{\uparrow} e_{\downarrow} \rangle $ as an order parameter.  

 \begin{figure}[t]
\begin{center}
\includegraphics[width=0.6\columnwidth]{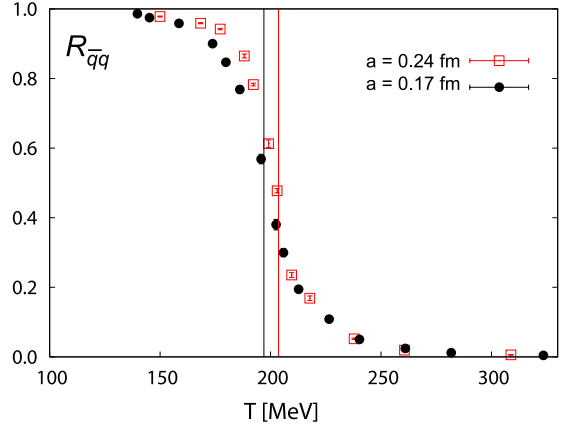}
\end{center}
\caption{Normalized chiral condensate 
${\cal R}_{\bar{q}q}=[\la \bar{\rm u}{\rm u} \ra \! - \! (\hat{m}/m_{\rm s})
 \la \bar{\rm s}{\rm s} \ra]/
[\la \bar{\rm u}{\rm u} \ra_0 \! - \! (\hat{m}/m_{\rm s}) \la \bar{\rm s}{\rm s} \ra_0]$
 as a function of $T$ for two different lattice spacings,
 $a= 0.24\ {\rm fm}$ and 
 $a= 0.17\ {\rm fm}$, calculated by the 
  lattice QCD simulations with dynamical u, d, s quarks in the 
  staggered fermion formalism.
   Vertical band in the middle indicates the 
  pseudo-critical temperature $T_{\rm pc}$  \cite{Cheng:2007jq}.}
\label{fig:qqbar-T-lat} 
\end{figure}
 
 To see that  $\langle \bar{q}q \rangle$ vanishes at extreme high $T$,  
 let us consider the QCD partition function at zero baryon density,
 $Z_{\rm QCD}  =  {\rm Tr} \left[  {\rm e}^{- \hat{H}_{\rm QCD}  /T} \right] 
  \equiv {\rm e}^{P(T)V/T} $, which leads to
$ \langle \bar{q}q \rangle 
   = - \frac{\partial P(T)}{\partial m_q}.$
 If we have a situation where  $T \gg \Lamqcd$, the system is  
  approximated by
   the Stefan-Boltzmann gas of free quarks and gluons
   because of the  asymptotic freedom.
%
  Since  the quark-gluon vertex does not change chirality (the transition
    between $\qL$ and $\qR$ is not allowed in perturbation theory)  and thus the 
     expectation value of $\bar{q}q= \bqL \qR+ \bqR \qL$ vanishes
     in any finite order of the perturbation as long as  $m_q=0$.
 
 As shown in Fig.\ref{fig:qqbar-T-lat},  lattice QCD Monte Carlo simulations
  at finite $T$ with zero baryon density 
   indeed indicate a sudden drop of the chiral condensate  
    around the pseudo-critical temperature determined from the 
    susceptibility peak \cite{Bazavov:2009zn, Aoki:2009sc},
 \beq
  T_{\rm pc} \simeq  (150-200)  \ {\rm MeV} = (1.7-2.3) \times 10^{12} \  {\rm K} .
 \eeq   
  Note that the phase transition 
  from the low $T$ phase (Nambu-Goldstone phase) to the high $T$ phase (quark-gluon
   plasma) is first order  in the flavor-SU(3)
    chiral limit ($m_{\rm u,d,s}=0$) and is
   second order  in the flavor-SU(2) chiral limit 
    ($m_{\rm u,d}=0, m_{\rm s}=\infty$)  \cite{Pisarski:1983ms}.
  On the other hand, in the presence of the finite 
  quark masses $m_{\rm u,d,s}$ acting as external fields,  the transition is  crossover
  as seen in Fig.\ref{fig:qqbar-T-lat}.
    For more details
    on the QCD thermodynamics, see the review \cite{DeTar:2009ef}. 

 The high temperature  quark-gluon plasma (QGP) is believed to be present in the 
  early universe  with its age younger than $10^{-5}$ sec.
  Attempts to create such extremely hot system by the 
   relativistic heavy-ion collisions  in the laboratory have been started
  from 2000 at  RHIC (Relativistic
 Heavy Ion Collider) in Brookhaven national laboratory and will be 
 pursued at LHC (Large Hadron Collider) in CERN.
  RHIC has already produced a plenty of data
 showing not only the evidence of QGP but also 
 strongly interacting characters of QGP \cite{Yagi:2005yb}. 
 (See the recent reviews 
 \cite{Muller:2007rs}
 on this  rapidly developing subject.)
 In the following,  we will focus more on QCD matter  with finite
  baryon density at low $T$. Such a system 
   may undergo successive quantum phase transitions which are
  relevant to the physics of  neutron stars and of
   possible quark stars.
   
\section{QCD matter at high baryon density}

 
 Soon after the discovery of the asymptotic freedom  of QCD,
   a possible transition from hadronic matter to quark matter
   in the core of the neutron stars
   has been pointed out \cite{Collins:1974ky}.
 A strange quark star
 entirely made of deconfined u, d, s quarks, yet undiscovered, 
 was also proposed \cite{Witten:1984rs} 
 as a modern version of the early idea of the quark star \cite{Itoh:1970uw}. 
 Although there has been no observational evidence of quark stars yet, 
 nature may be strange enough to accommodate such compact object 
 in our universe and  we should  prepare for the future discovery. 

\begin{figure}[t]
\begin{center}
\includegraphics[width=0.7\columnwidth]{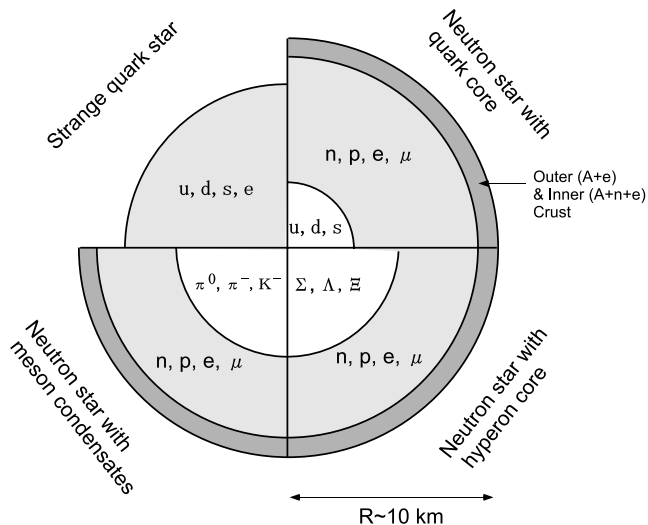}
 \end{center}
\caption{Possible internal structures and compositions of  
 four different types of compact stars \cite{Yagi:2005yb}. }
\label{fig:ns-cross}
\end{figure}

Shown in Fig.\ref{fig:ns-cross} is a schematic
 view of various forms of compact stars, from 
 the neutron star to the quark star.
 Typical radius of the neutron star is about $ 10 \ {\rm km}$,
 while  its mass 
 is comparable to the solar mass 
  $M_{\odot} (\simeq 2 \times 10^{30}\ {\rm kg})$.
  Since neutrons cannot be bound 
  by the strong interaction only, the presence of
    gravitational force is essential to hold the neutron star.
 This implies that the radius increases as the mass decreases.
  The masses of the neutron stars in the 
  binary systems are centered
   around $1.35 M_{\odot}$. 
 Observed upper limit of the surface temperature of neutron stars
  is less than $10^9$ K after 1 year from their births.
   In the early stage, the cooling  occurs through
    the neutrino emissions, while in the  later stages it is
     dominated
     by surface photon emissions. 
     
         For pulsars (rotating neutron stars), the 
   measured rotational frequency  is ranged from 
     milli second to several seconds.  
  The surface magnetic field  
  is typically $10^{12} $ gauss for 
  ordinary pulsars with  rotational period $P\sim \ 1\ {\rm s}$ 
  and $dP/dt\sim 10^{-15}$. There are
   also  stars with much larger (smaller) $dP/dt$ and
    larger (smaller) magnetic field  $\sim 10^{15} (10^9)$ gauss.
   Sudden spin up of the rotation associated with a subsequent
   relaxation to the normal rotation
    has been observed and  is called the glitch.  This
  phenomena should be 
    related to the internal structure of  
   neutron stars, in particular the superfluidity of the 
   neutron liquid.
  
  The outer crust of the neutron stars is a solid composed of 
  heavy nuclei forming a Coulomb lattice in the 
   sea of degenerate electrons.
   As the pressure and the density increase toward the 
   inner region, electrons tend to be captured
   by nuclei and at the same time neutrons drip 
    out from the nuclei, so that the system is
     composed of neutron-rich heavy nuclei 
      in the Fermi sea of the neutrons and electrons.
  Eventually, the nuclei dissociate into neutron
   liquid and the system becomes a degenerate
    Fermi system composed of superfluid
    neutrons together with  a small fraction of
     superconducting protons and  normal electrons.

  When the baryon number density ($\rho$) of the core of the neutron stars
  exceeds
  a few times of the central density of heavy atomic nuclei  
  ($\rho_{0}= 0.16\ {\rm fm}^{-3} = 0.16 \times 10^{39}\ {\rm cm}^{-3}$),
   one may expect
  exotic components  such as the hyperons (baryons with s quarks),
  Bose-Einstein condensates of pions and kaons, and the 
    deconfined quark matter, which can contribute to the 
    acceleration of the neutron-star cooling.
 For more details of the physics of high density matter 
 and  compact stars, see e.g. \cite{Heiselberg:2000dn, Lattimer:2006xb}.

\begin{figure}[t]
\begin{center}
\includegraphics[width=0.8\columnwidth]{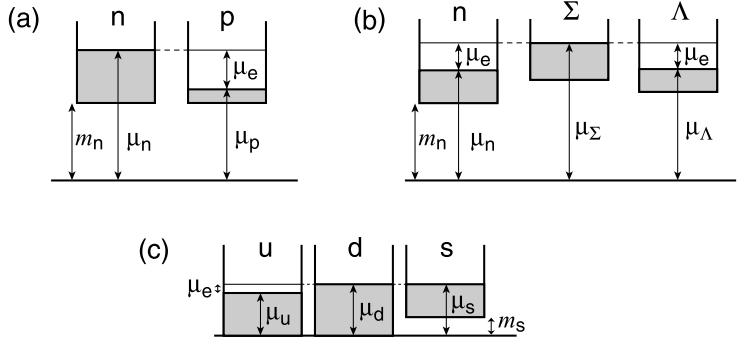}
 \end{center}
\caption{ 
Compositions of matters under chemical equilibrium and 
 charge neutrality conditions in the Fermi gas model.
 Shaded areas show the occupied states \cite{Yagi:2005yb}.
  (a) The neutron-star matter
  with  n, p, and e$^-$,
  (b) the hyperon matter with n, $\Sigma^-$ and $\Lambda$, 
  (c) the u-d-s quark matter with finite strange quark mass $m_{\rm s}$.
  }
\label{fig:fermi}
\end{figure}
 
 \subsection{Neutron-star matter and hyperonic matter}
 
 Although the neutron star is  mainly composed of the degenerate neutrons,
 other species are also present as a result of the
  chemical equilibrium conditions.
 Indeed,  the matter made of only neutrons
 is unstable against the 
  $\beta$-decay, ${\rm n} \rightarrow {\rm p} + {\rm e}^- + \bar{\nu}_{\rm e}$.
   After the decay, the electron-neutrino leaves the star
    without much interactions if the neutron star is cold enough.  
             On the other hand, the protons and the electrons
    remain in the star and form degenerate Fermi liquid together
     with the neutrons.
    The equilibrium configuration of n, p and ${\rm e}^-$,
     which we call the standard  
     neutron-star matter (see Fig.\ref{fig:fermi}(a)),   is
     determined by the three conditions: chemical equilibrium,
    charge neutrality and the baryon-number conservation:
$ \mu_{\rm n}  =  \mu_{\rm p} + \mu_{\rm e}$,
$\rho_{\rm p} =  \rho_{\rm e}$, 
$ \rho = \rho_{\rm n} + \rho_{\rm p}$, where
 $\rho_i$ denotes the  number density of $i$-species.
 
 If we assume the non-interacting  degenerate fermions for simplicity,
  it is easy to find the analytic solution of the above conditions; 
     e.g. the proton fraction in a neutron star
  for a given baryon number density $\rho$ reads
  $
  {\rho_{\rm p}}/{\rho_{\rm n}}
   \simeq \frac{1}{8} \left[ 1+ \left({m_{\rm n}^3}/{3\pi^2 \rho_{\rm n}}
    \right)^{2/3}  \right]^{-3/2}$.
 This is a monotonically increasing function of $\rho_{\rm n}$
   and approaches to the asymptotic limit $1/8$  from below.
  As the neutron density further increases and the electron chemical potential
 exceeds the muon mass, $\mu_{\rm e} > m_{\mu}=105$ MeV,  
  the system composed of    n, p, ${\rm e}^-$ and $\mu^-$ is realized.
  
 As the baryon density increases further,  
 hyperons 
 enter into the game. This is because the 
  Fermi energy of the neutron exceeds the threshold
  of the neutron-decay into hyperons. See Fig.\ref{fig:fermi}(b). 
  Hyperons such as $\Sigma^-$ and $\Lambda$ may appear for $\rho > (2-3) \rho_{0}$.
   Which hyperon appears first depends on the still uncertain
   hyperon-nucleon  interactions.
 
\subsection{Quark matter}
 
As the baryon number density $\rho$ of the system exceeds $ (3-5) \rho_{0}$, the 
 neutrons, protons and hyperons start to percolate each other due to their
 finite sizes \cite{Baym:1979}.  The quark number density $\rho_q$ for each flavor
  is related to $\rho$ as $\rho_q = (N_c/N_f) \rho$ with $N_c=3$ being the 
   number of colors and $N_f$ the active number of flavors so that the critical
    quark chemical potential, above which the percolation to quark matter
     takes place, is estimated as
 \beq
\mu_{\rm c} = \left( \pi^2 \rho_q \right)^{1/3}
\simeq (380 - 450) \ {\rm MeV}.
\eeq
Here we have assumed a
   non-interacting  and degenerate quark matter composed of 
    massless u and d quarks.
   
 Let us now consider the quark matter composed of only
   u, d and  e$^-$. 
The condition of chemical equilibrium $({\rm d} \leftrightarrow {\rm u}+{\rm e}^-)$, charge
 neutrality and the  baryon
 number conservation read
$\mu_{\rm d}          =  \mu_{\rm u} + \mu_{\rm e}$,   
$\frac{2}{3} \rho_{\rm u} - \frac{1}{3} \rho_{\rm d} - \rho_{\rm e} = 0$,
 and $\frac{1}{3} (\rho_{\rm u} + \rho_{\rm d}) = \rho $.
The factors $2/3$ and $-1/3$  originate from the electric charges of the 
quarks and $1/3$ from the baryon number of a quark.
 If we assume non-interacting  quarks at high density
 ($\mu_q \gg m_q$),    one immediately finds
$\mu_{\rm u} \simeq 0.80 \ \mu_{\rm d} $.
Thus the Fermi energy of the d quark is slightly higher 
 than that of the u quark, which is
   different from the situation of 
 neutron matter where n and p have quite a 
 different Fermi  energies due to non-relativistic kinematics as shown
  in Fig.\ref{fig:fermi}(a). 
  
 If the quark matter is composed of 
   u, d, s  and  e$^-$,
  the chemical equilibration
 is achieved through the processes, 
 ${\rm d} \leftrightarrow {\rm u} + {\rm e}^-$, 
 ${\rm s} \leftrightarrow {\rm u} + {\rm e}^- $,
 and ${d} + {u} \leftrightarrow  {u} + {s}$. 
 Then the  equilibrium conditions read
$\mu_{\rm d}  =  \mu_{\rm u} + \mu_{\rm e}$, 
$\mu_{\rm s} = \mu_{\rm d}$,
$ - \frac{1}{3} (\rho_{\rm d} + \rho_{\rm s}) + \frac{2}{3} \rho_{\rm u} 
- \rho_{\rm e} =0$,
 and $\frac{1}{3} (\rho_{\rm u} + \rho_{\rm d} + \rho_{\rm s}) = \rho$.
 For  $\mu_{q} \gg m_q$, 
 they lead to 
$\mu_{u} = \mu_{d} = \mu_{s}$ and $\mu_{\rm e} =0$.
Namely the massless u-d-s  quark matter is charge neutral by itself without electrons.
 If we have finite $m_{\rm s}$, then
  $\rho_{\rm s}$ is reduced relative to $\rho_{\rm u,d}$
   and the electrons become necessary  to make  the system charge neutral,
   as shown in Fig.\ref{fig:fermi}(c).

\section{Superfluidity in neutron-star matter}
\label{sec:SFN}

  If there exists  attractive channel between the 
  fermions near the Fermi surface, the system
  undergoes a transition to superfluidity or superconductivity 
  in three spatial dimensions \cite{Shankar:1993pf}.  
  This is indeed the case in the neutron-star matter 
  where the attraction between the protons due to spin-independent nuclear 
   force     in the $(S,L,J)$=(spin, orbital angular momentum,
   total angular momentum)=(0,0,0) channel
   leads to
   the condensation of  $^{2S+1}L_J$=$^1{\rm S}_0$ Cooper pairs
    (the proton superconductivity) and 
  the attraction between the neutrons due to spin-orbit nuclear force  
   in the $(S, L, J)$=(1,1,2) channel  leads to
   the condensation of $^3{\rm P}_2$ Cooper
    pairs (the neutron superfluidity) 
   \cite{Tamagaki:1970}. 

 Other than the superfluidity and superconductivity of the nucleons,
  the condensation of pions ($\pi^{0}$ and $\pi^-$)
  and that of kaons (${\rm K}^-$)  have been  
  studied extensively. 
  For more details on nucleon  superfluidity, meson condensation
     and its   implication to the physics of 
   compact stars, see the reviews \cite{KM93, Le96, Dean:2002zx}.

\section{Color superconductivity in quark matter}
\label{sec:CSC}

 The quark matter exhibits
 the  color superconductivity (CSC) which originates from the 
  formation of Cooper pairs of quarks near the Fermi surface (see the recent 
  reviews
    \cite{Rajagopal:2000wf,Alford:2007xm} 
  and references therein).
  Dominant attractive interaction responsible for the quark$-$quark 
 pairing at high density is the  color-magnetic interaction mediated by the gluon.
 
  There are some characteristic differences between CSC and  the 
  standard BCS-type superconductivity:
 \begin{enumerate}
  \item[(i)] 
 The quark matter at high density is  a
  relativistic system where
  the quark chemical potential $\mu$ is comparable or larger than the quark mass $m_q$.
  In such a case, the velocities of quarks near the Fermi surface are close to light velocity and the  
  magnetic interaction is not any more suppressed in comparison to the electric interaction.
 \item[(ii)]  The color-magnetic interaction
  is screened  only dynamically by Landau-damping, while the color-electric
   interaction is Debye screened as usual. Therefore, the 
   collinear quark$-$quark scattering on the Fermi surface is 
    dominated by the color-magnetic interaction, which leads
    to an unconventional form of the fermion gap 
     $ \Delta \propto \mu\: e^{-c/\sqrt{\als}} $ \cite{Son:1998uk}.
 Because of this non-BCS form  where 
    the coupling strength enters as $\sqrt{\als}$ instead of $\als$,
   $\Delta/\mu$  becomes sizable even in the weak coupling. 
 \item[(iii)] Due to color and flavor indices of the
    quarks, the CSC gap acquires color-flavor matrix structure, which
 leads to various different phases depending on $\mu$ and $T$;
 they include the 2SC 
  (2-flavor color superconducting) phase, the
    CFL (color-flavor locked) phase, 
     the FFLO phase, the crystalline phase, and so on 
 as reviewed in \cite{Alford:2007xm}. 
 \end{enumerate}

\subsection{The gap equation}
\label{subsec:gap-eq}

 Let us illustrate the role of the above color-magnetic interaction 
 by considering a simplified situation where  
quark matter is composed of only massless u and d quarks with equal Fermi energies 
at $T=0$.
  In this case, the u quark with red-color
  and the d quark with green-color in flavor-singlet and 
  color anti-triplet combinations are paired, while all the others quarks
 are unpaired  \cite{Bailin:1983bm}.  This is called the 2SC phase.
 %
 
  Using the standard Nambu-Gor'kov field  $\Psi=(q,\bar{q}^t)^t$,  
 the Schwinger-Dyson equation for the 
  quark self-energy $\Sigma$ in the ladder approximation is written as
\beq
\label{eq:gap-equation}
  \Sigma(k)&=&-i \int\frac{d^4p}{(2\pi)^4} \ g^2(p,k) \ 
 \Gamma_{a}^\mu S(p)%
  \Gamma_{b}^\nu D_{\mu\nu}^{ab}(p-k),\label{DSeq}
\eeq
where $D^{ab}_{\mu\nu}$ is the in-medium gluon propagator and
 $\Gamma^{a}_\mu$ is the bare quark-gluon vertex.
   The off-diagonal (anomalous) 
  component of $\Sigma$  in the Nambu-Gor'kov space   is directly 
  related to  the gap function  $\Delta(k) = \Delta_+(k) \Lambda_+ + \Delta_-(k) \Lambda_-$,
  where $\Lambda_{\pm}$ are the projection operators to the positive and negative energy
   states. Therefore, $\Delta_+(k)$ and $\Delta_-(k)$ are interpreted as the 
  quark gap and anti-quark gap, respectively. 
  As for  $g(q,k)$,
  the  Higashijima-Miransky ansatz 
  (a momentum-dependent QCD coupling with phenomenological
   infrared regulator) may be adopted   \cite{Higashijima:1991de}.
 As for the  gluon propagator, we take the screened propagator   
in the Landau gauge, 
\beq
  D^{ab}_{\mu\nu}(k)=\biggl(-\frac{P^{\rm L}_{\mu\nu}}{\bfm{k}^2+m_{\rm D}^2}
-\frac{P^{\rm T}_{\mu\nu}}{\bfm{k}^2+i \frac{\pi}{4} m_{\rm D}^2 |k_0|/|\bfm{k}|}
\biggl)\delta^{ab},
 \label{propagator}
\eeq
where $P^{\rm L,T}_{\mu\nu}$ are the longitudinal and  transverse 
projection operators. 
 The longitudinal (electric)  part of the propagator has static 
 screening by the Debye mass
 $m_{\rm D}^2 = (1/\pi^2) g^2 \mu^2 $, so that the static interaction between quarks in
  the coordinate space is the
 Yukawa-type short range potential.
 On the other hand, the
 transverse (magnetic) part has only  
 dynamical screening due to Landau damping, so that 
  the static interaction between quarks in
  the coordinate space is the
 Coulomb-type long range potential.

 In the high density limit (or equivalently the weak 
 coupling limit due to asymptotic freedom), 
 only the interactions near the Fermi surface 
 become relevant and the gap equation can be  simplified
 to give the form \cite{Son:1998uk}
\beq
     \Delta_+(|\bfm{k}|=\mu)  \cong  2 b  \ \mu\ %
        e^{-(3\pi^2/\sqrt{2})/g(\mu)},
\label{sol-LD}
\eeq 
with  $b =  256 \pi^4/g^5(\mu)$ and $g(\mu)=\sqrt{4\pi \als(\kappa=\mu)}$
 \cite{Alford:2007xm}.
The characteristic form of the gap, $e^{-c/g}$, in 
  Eq.(\ref{sol-LD}) originates from the long-range color-magnetic
  interaction  in Eq.(\ref{propagator})
    and  is different from  the BCS form $e^{-c/g^2}$ as we have mentioned.  
   
\begin{figure}[t]
\begin{center}
\includegraphics[width=0.46\columnwidth]{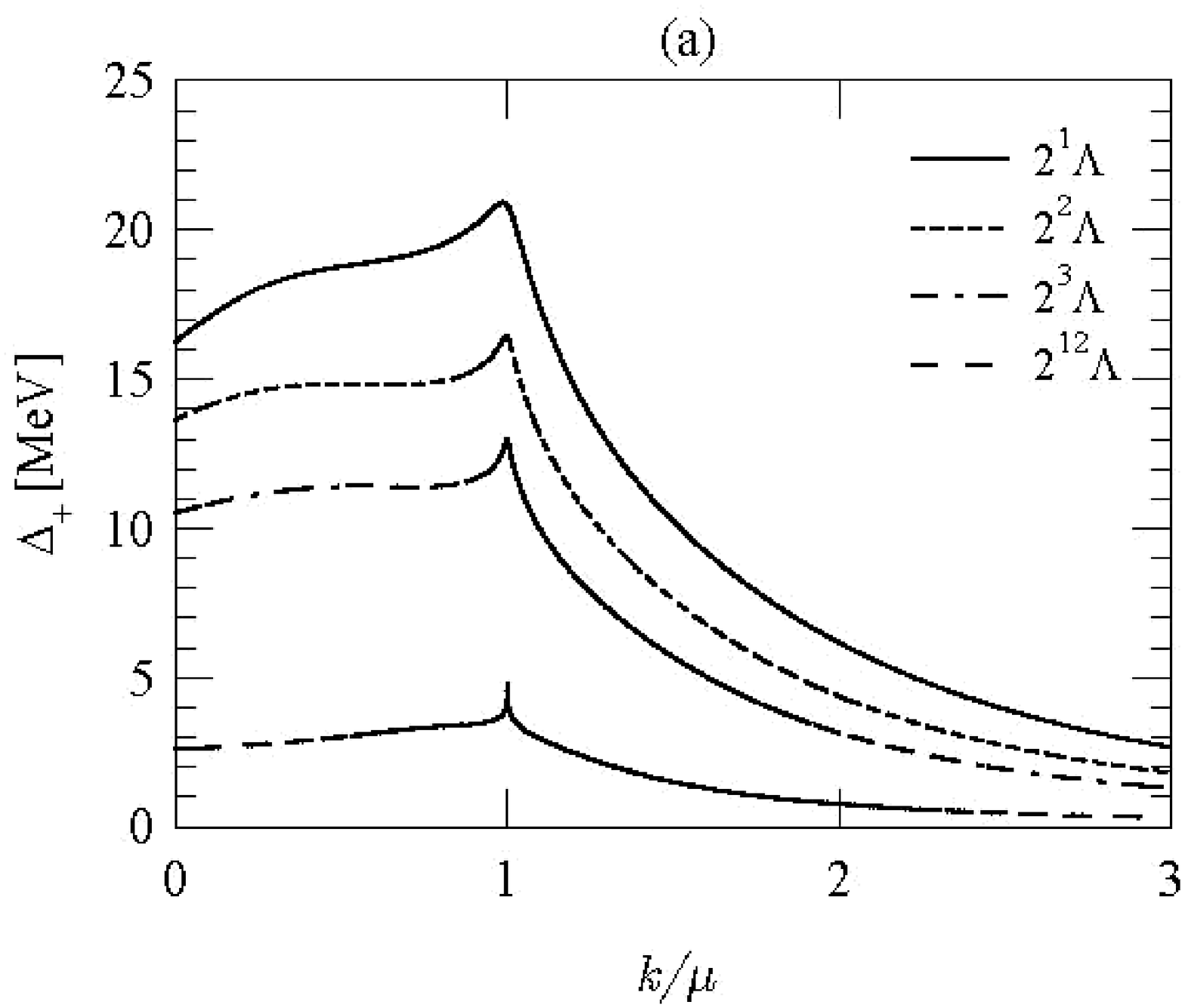}
\hspace{0.25cm}
\includegraphics[width=0.46\columnwidth]{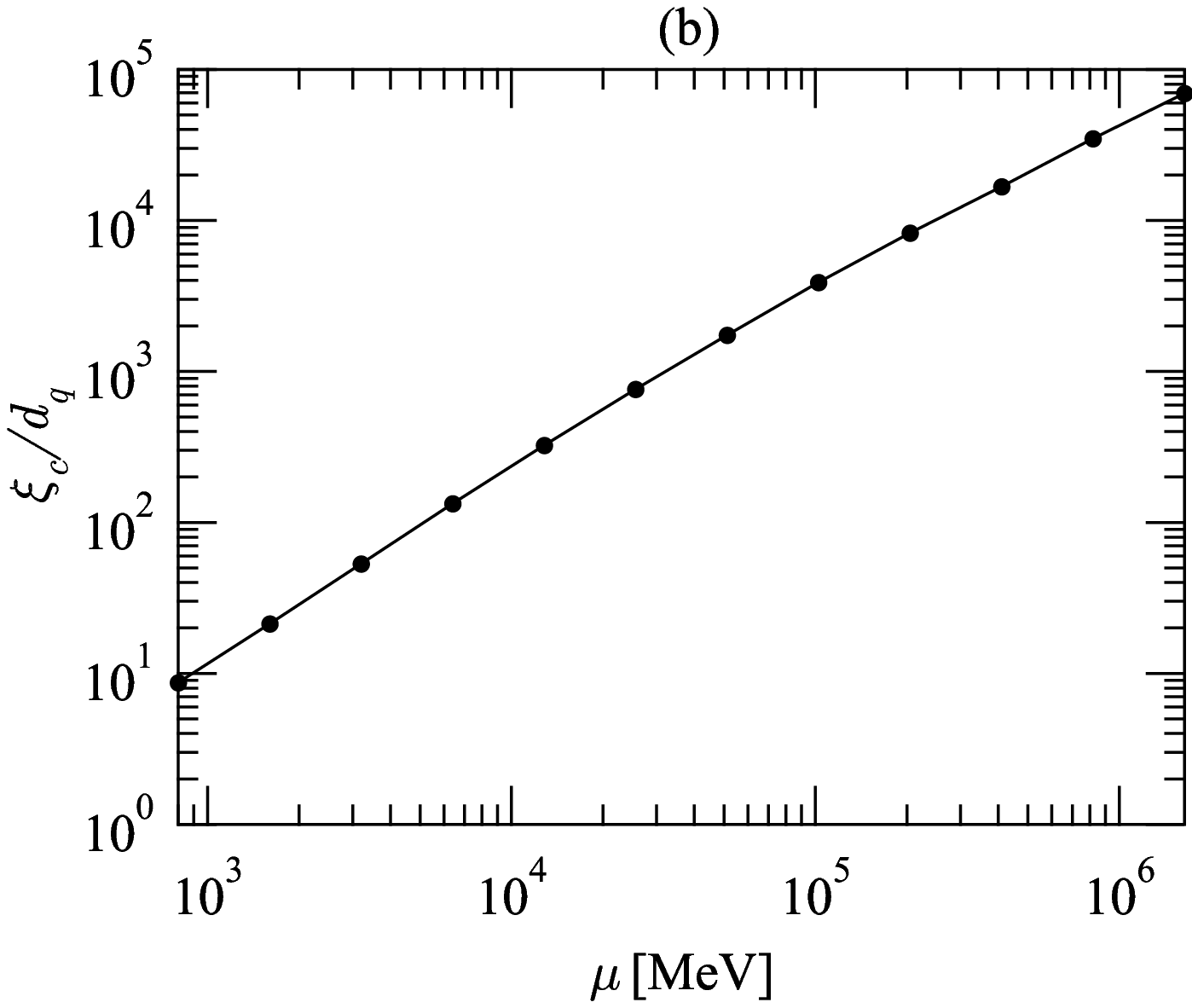}
\end{center}
  \caption{%
   (a) The quark gap  $\Delta_+(\bfm{k})$ as a function of $|\bfm{k}|/\mu$ for various 
   quark chemical potentials,
 $\mu =2^n \times \Lambda$ with 
     $\Lambda=400\ $ MeV and    $n=1,2,3,12$, obtained 
   by solving the gap equation Eq.(\ref{eq:gap-equation}) numerically \cite{Abuki:2001be}.
   (b) Ratio of the coherence length $\xi_{\rm c}$  and the
          average inter-quark distance $d_q$ as a function of $\mu$.}
  \label{gap-sol}
\end{figure}

At low densities,
  sizable diffusion of the Fermi surface occurs, and the
  weak-coupling approximation in the asymptotic high density leading to Eq.(\ref{sol-LD})
  is not justified.
 Therefore, we solve
  the gap equation  Eq.(\ref{DSeq}) numerically
  with both magnetic and electric interactions \cite{Abuki:2001be}.
 In Fig.\ref{gap-sol}(a), we show such numerical solution of 
  $\Delta_+(k)$  as a function of  $|\bfm{k}|/\mu$ for wide range of density.
  The figure shows that (i)  the ratio of the gap and the quark chemical potential
   $\Delta_+/\mu$ can be sizable in magnitude of order 0.03 even
   for baryon density as high as $\rho \sim 10 \rho_0$, and that
  (ii) the gap  function $\Delta_+(\bfm{k})$ has non-trivial  momentum dependence
 with a peak near the Fermi surface. 
   
\subsection{Tightly bound Cooper pairs}

 The color superconductivity is a strongly coupled system 
    partly due to the large value of  
  coupling constant $\als (\kappa=\mu)$ and
   partly due to  long-range nature of the color-magnetic interaction.  
To clarify this point further,
  let us consider
  the coherence length
 $\xi_{\rm c}$  defined  as a root-mean-square radius
 of the Cooper-pair wave function,
$\varphi_+(\br) \propto \langle q(\br) q ({\bf 0}) \rangle \propto
  \int \frac{d^3 \bks}{(2\pi)^3} 
  e^{i \bks \cdot \brs } \frac{\Delta_+(\bks)}{2\sqrt{(|\bks|-\mu)^2+\Delta_+^2(\bks)}}$.

Shown in Fig.~\ref{gap-sol}(b) is the ratio of $\xi_{\rm c}$
  and the  average inter-quark distance 
 $d_{q}=({\pi^2}/{2})^{1/3}{\mu}^{-1}$ in the u-d quark matter.
  At high density, this ratio is very large (about $10^5$ at $\mu\sim 10^6$ MeV), while
  at low densities, the ratio becomes small (about $10$ at $\mu=800$ MeV) and 
   may even become less than 1  at lower densities.  
  This situation is quite similar to the BCS-BEC crossover phenomenon 
  \cite{Leggett:1980} 
which was
  recently observed in ultracold atomic systems \cite{Regal:2004}: 
  The result here suggests that the quark matter possibly realized in
    the core of neutron stars $(\mu \sim 400 $ MeV) may be rather like the BEC of tightly bound
    Cooper pairs.  For further studies of the BCS-BEC crossover in the relativistic system,
     see e.g. \cite{Nishida:2005ds}.

\section{QCD phase structure}

\begin{figure}[t]
\begin{center}
\includegraphics[width=0.7\columnwidth]{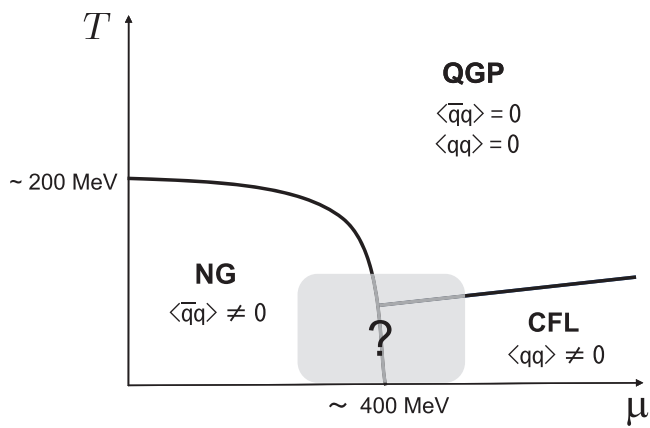}
\caption{
{\small The
three basic phases of QCD in the  $T$-$\mu$ plane 
for flavor-SU(3) chiral limit 
.
 }}
\label{fig:T-mu}
\end{center}
\end{figure}

The continuous QCD symmetry in the flavor-SU(3) chiral limit ($m_{\rm u,d,s}=0$),
Eq.(\ref{eq:GG-symmetry}), exhibits various symmetry breaking patterns depending on the 
 temperature $T$ and the quark chemical potential $\mu$.
 In Table \ref{tab:SSB-pattern}, three examples of the 
 symmetry realizations are shown;
the QGP (quark-gluon plasma) 
phase, the NG (Nambu-Goldstone) phase and  the CFL (color-flavor locked) phase.
(CFL is one of the color superconducting phases to be discussed later in more details.)
 Each phase would appear
  in the $T-\mu$ phase diagram
 as illustrated  in Fig.\ref{fig:T-mu}.  
 In the intermediate values of $\mu \sim 400$ MeV at low $T$,
  a variety of quantum phases have been proposed \cite{KM93,Alford:2007xm}.

   As we have discussed in Sec.\ref{subsec:DBCS}, the NG phase is characterized by the 
 dynamical breaking of chiral symmetry due to  nonvanishing 
 chiral condensate $\langle \bar{q}q \rangle$.
 On the other hand, in the color superconductivity 
  at high chemical potential, the diquark condensate $\langle qq \rangle$ 
  is formed as discussed in Sec.\ref{sec:CSC}.
In the presence of these condensates,
 the light quarks and anti-quarks acquire  the Dirac type mass $M$ 
 and the Majorana type mass $\Delta_{\pm}$, which  
  leads to the relativistic quasi-particle spectrum of a quark 
  near the Fermi surface,
\begin{eqnarray}
 \omega (\bp) = \sqrt{ (\sqrt{\bp^2 + M^2} - \mu)^2+|\Delta_+|^2 } .
\label{eq:DM-disp}
\end{eqnarray}  

\begin{table}[t]
 \caption{\small Symmetry breaking patterns of QCD
  in the flavor-SU(3) chiral limit ($m_{\rm u,d,s}=0$).
 The QGP phase, the NG phase, and the CFL phase denote 
the quark-gluon plasma phase, the Nambu-Goldstone phase, 
and  the color-flavor locked phase, respectively.}
  \begin{center}
  \begin{tabular}{|c|c|c|}
    \hline 
 phase  & region & unbroken continuous symmetries \\
   \hline \hline 
 QGP phase &
 $\frac{T}{\Lamqcd} \gg 1$  &
 ${\rm SU(3)}_{\rm C} \otimes {\rm SU(3)}_{\rm L} \otimes {\rm SU(3)}_{\rm R} \otimes {\rm U(1)}_{\rm B}$ \\
\hline
NG phase  &
 $\frac{T}{\Lamqcd} \ll 1 $, $\frac{\mu}{\Lamqcd} \ll 1$   &  
 ${\rm SU(3)}_{\rm C} \otimes {\rm SU(3)}_{\rm L+R} \otimes {\rm U(1)}_{\rm B}$ \\ 
 \hline
CFL phase &
 $ \frac{T}{\Lamqcd} \ll 1$,  $\frac{\mu}{\Lamqcd} \gg 1$ &
 $ {\rm SU(3)}_{\rm C+L+R} $   \\
    \hline 
  \end{tabular}
 \end{center}
\label{tab:SSB-pattern}
\end{table}
   
\subsection{Ginzburg-Landau potential for hot/dense QCD}
\label{subsec:GL}

 In this subsection, we focus our attention on the QCD phase structure
   at intermediate value of the chemical potential $(\mu \sim 400$ MeV)
   where there would be an 
  interplay between 
  the quark$-$anti-quark pairing characterized by the
   chiral condensate $\langle \bar{q}q \rangle$
    and the quark$-$quark pairing characterized by the
 diquark condensate $\langle qq \rangle$ \cite{Kitazawa:2002bc,Hatsuda:2006ps}. 
 Study of this region  is not
  only important to understand the quantum phase transition to quark matter
   in the deep interior of the neutron stars, 
  but also interesting in relation to similar phenomena in other
 systems such as  the interplay between magnetically ordered phases and metallic
 superconductivity \cite{Sigrist:1991}  and that between
 superfluidity and magnetism in ultracold atoms \cite{Cherng:2007}.

 To analyze such interplay in QCD in a model-independent manner, 
 let us construct a Ginzburg-Landau (GL) potential  $\Omega$
 on the basis of the 
  QCD symmetry   Eq.(\ref{eq:GG-symmetry}) as
$ \Omega(\Phi,\dL,\dR)
  = \Omega_{\chi}(\Phi)+\Omega_d(\dL,\dR) + \Omega_{\chi d}(\Phi,\dL,\dR)$.
 Here the chiral field $\Phi$, which has a 3$\times$3 matrix structure in the flavor space,
 is defined  with the 
 transformation property under Eq.(\ref{eq:quark-rotation}) as,
\beq
\label{eq:phi-def}
   \Phi_{ij} \equiv \langle [{\bqR}]_{a}^j [\qL]_{a}^i \rangle , \ \ \ 
 \Phi \rightarrow {\rm e}^{-2i \theta_{\rm A}} \ V_{\rm L}  \Phi  V_{\rm R}^{\dagger} .
\eeq
 On the other hand, the diquark field $\dL$,
  which has a 3$\times$3 matrix structure in the flavor-color space,
  is defined  with the 
 transformation property as 
\beq
\label{phi-dl}
   [\dL^{\dagger}]_{ai} \equiv
    \epsilon_{ijk} \epsilon_{abc} \langle [\qL]_{b}^j  C  [\qL]_{c}^k \rangle  , \ \ \ 
   \dL \to e^{2i \theta_{\rm A}}  e^{2i \theta_{\rm B}}\  V_{\rm L}  \dL  V_{\rm C}^{t},
\eeq
where $C=i \gamma^2 \gamma^0$ is the charge conjugation matrix.
 Similar definition holds for $\dR$ too.
 By definition, the 3$\times$3 matrix
$[ \dLR ]_{ia}$ belongs to the  fundamental representation of
${\rm SU(3)}_{\rm C}$ and ${\rm SU(3)}_{\rm L(R)}$.   

   Most general form of the GL potential which is invariant under ${\cal G}$ in
   Eq.(\ref{eq:GG-symmetry}), written in terms of 
   the chiral field
   up to  ${\cal O}(\Phi^4)$,   reads \cite{Pisarski:1983ms},
\beq
   \label{eq:GL-chi}
\! \! \! \!  \! \! \!   \! \! \!  
 \Omega_{\chi} = \frac{a_0}{2} {\rm Tr}  \  \Phi^{\dagger} \Phi
   + \frac{b_1}{4!} \left( {\rm Tr} \ \Phi^{\dagger} \Phi    \right)^2
   + \frac{b_2}{4!} {\rm Tr} \left( \Phi^{\dagger} \Phi    \right)^2
     - \frac{c_0}{2} \left( {\rm det} \Phi + {\rm det} \Phi^{\dagger} \right),
\eeq
where ``Tr" and ``det" are taken over the flavor indices, $i$ and $j$.
  The first three terms in the right hand side are 
  invariant under ${\cal G} \otimes {\rm U(1)}_{\rm A}$, while the  
  last term represents the axial anomaly which  breaks  ${\rm U(1)}_{\rm A}$ 
  down to ${\rm Z(6)}_{\rm A}$.  The potential 
$\Omega_{\chi}$ is bounded from below for
$b_1+b_2/3>0$ and $b_2>0$.  If these conditions are not satisfied, we need to
introduce terms in ${\cal O}(\Phi^6)$ to stabilize the potential,  a situation
we will indeed encounter.  We assume $c_0$ to be positive so that the
chiral condensate at low temperature is positive.
Also, we assume that $a_0$ changes its sign at a certain
temperature to drive the chiral phase transition.

   Most general form of the GL potential which is invariant under ${\cal G}$,
    written in terms of 
    the diquark field
      up to ${\cal O}(d^4)$,     reads \cite{Iida:2000ha, Iida:2004cj},
\beq
   \label{eq:GL-d}
   \Omega_{d} &=&
     \alpha_0 \ {\rm Tr} [\dL^{\ } \dL^{\dagger} +\dR^{\ }  \dR^{\dagger} ] \nonumber \\
   & &+ \beta_{1} \left(  [ {\rm Tr}(\dL^{\ } \dL^{\dagger})]^2
                        + [ {\rm Tr}(\dR^{\ } \dR^{\dagger})]^2    \right)
    + \beta_{2}  \left(  {\rm Tr} [(\dL^{\ } \dL^{\dagger})^2]
                        + {\rm Tr} [(\dR^{\ } \dR^{\dagger})^2]    \right) \nonumber \\
 & &+ \beta_3 \ {\rm Tr} [(\dR^{\ } \dL^{\dagger})(\dL^{\ }\dR^{\dagger})]
 +  \beta_4 \ {\rm Tr} (\dL^{\ }\dL^{\dagger}){\rm Tr}(\dR^{\ }\dR^{\dagger}).
\eeq
The transition from the normal state to color superconductivity is driven by $\alpha_0$
changing sign.  Unlike det\ $\Phi$ in $\Omega_{\chi}$,
 terms such as ${\rm det}\ \dLR$ are not
allowed in $\Omega_{d}$, since det\ $\dLR$ carries baryon number and is not
 invariant under U(1)$_{\rm B}$.

    Finally, the interaction potential which is invariant under ${\cal G}$,
     written in terms of both chiral and diquark fields to
 fourth order, reads \cite{Hatsuda:2006ps,Yamamoto:2007ah},
\beq
\label{eq:GL-coup}
   \Omega_{\chi d}
   &= & \gamma_{1} \ {\rm Tr} [  (\dR^{\ } \dL^{\dagger})\Phi
                    + (\dL^{\ } \dR^{\dagger})\Phi^{\dagger}] \nonumber \\
  & & + \lambda_{1} \ {\rm Tr} [(\dL^{\ } \dL^{\dagger})\Phi \Phi^{\dagger}
                  +(\dR^{\ } \dR^{\dagger})\Phi^{\dagger}\Phi ] 
  +\lambda_{2} \ {\rm Tr} [\dL^{\ } \dL^{\dagger} + \dR^{\ } \dR^{\dagger}]
                      \cdot {\rm Tr} [\Phi^{\dagger}\Phi ] \nonumber \\
  & & + \lambda_{3} \left( {\rm det} \Phi \cdot
                    {\rm Tr}[(\dL^{\ } \dR^{\dagger}) \Phi^{-1}] + h.c. \right) .
\eeq
The  term with the coefficient $  \gamma_1$ 
originates from the axial anomaly which imposes the sign of $\gamma_1$
 in Eq.(\ref{eq:GL-coup})
 and that of $c_0$ in Eq.(\ref{eq:GL-chi}) being the same.
 
    Equations (\ref{eq:GL-chi}), (\ref{eq:GL-d}) and (\ref{eq:GL-coup})
constitute the most general form of the GL potential under the conditions
that the phase transition is not strongly first order (i.e., the magnitudes of
$\Phi, \dLR$ are sufficiently smaller than those at zero temperature) and
that the condensed phases are spatially homogeneous.
 To proceed analytically for the flavor-SU(3) chiral limit, we
restrict ourselves to maximally symmetric condensates of the form:
\beq
\label{DIA ansatz}
 \Phi ={\rm diag}(\sigma, \sigma,\sigma), \ \ \ 
 \label{CFL ansatz} \dL =-\dR={\rm diag}(d,d,d),
\eeq
where $\sigma$ and $d$ are assumed to be real and spatially uniform. 
 We have chosen the
relative sign between $\dL$ and $\dR$ in Eq.~(\ref{CFL ansatz}) so that
 the ground state has positive parity, as is indeed favored by the axial
anomaly together with finite quark masses.  
 The above ansatz for the  diquark condensate
 has residual symmetry ${\rm SU(3)}_{\rm C+L+R} \otimes {\rm Z(2)}$ and 
 is called the color-flavor locking (CFL) because of its symmetry realization
 \cite{Alford:1998mk}.  (Note that Z(2) corresponds to
 the reflection, $\qLR \rightarrow - \qLR$.)
 
The reduced GL potential with Eq.(\ref{DIA ansatz}) is
\beq
    \label{eq:nf3-model}
\! \! \! \! \! \! \! \! \! \! \!  \! \! \! \! 
     \Omega_{\rm 3F}
    = \left( \frac{a}{2} \sigma^2 - \frac{c}{3} \sigma^3 +
    \frac{b}{4}\sigma^4
    + \frac{f}{6}\sigma^6 \right)
    + \left( \frac{\alpha}{2} d^2 + \frac{\beta}{4} d^4 \right)
     - {\gamma} d^2 \sigma +  {\lambda} d^2 \sigma^2.
\eeq
 Here the axial anomaly leads to $c >0$ and $ \gamma >0$, while
  microscopic calculation based on the Nambu$-$Jona-Lasinio model as well as 
  the weak-coupling QCD suggests that
  $\lambda$ is  positive and plays a minor role in comparison to
   $\gamma$   \cite{Yamamoto:2007ah}. Note that
 we have introduced $f$-term ($f>0$) in case that $b$ becomes negative.
  This system can have four phases with the following dynamical breaking patterns 
  of continuous symmetries;
\beq
    {\rm QGP\  phase}      &: &  \sigma=0, d=0 \nonumber \\
     {\rm NG\ phase}        &: &  \sigma\neq 0, d= 0
 \ :\ {\cal G} \to {\rm SU(3)}_{\rm C} \otimes {\rm SU(3)}_{\rm L+R} \otimes {\rm U(1)}_{\rm B} \nonumber \\
    {\rm CFL\  phase}      &: &  \sigma=0, d\neq 0 
    \ :\ {\cal G} \to {\rm SU(3)}_{\rm C+L+R}  \nonumber \\
    {\rm COE \ phase} &: &  \sigma\neq 0, d \neq 0
 \ :\ {\cal G} \to {\rm SU(3)}_{\rm C+L+R}  .
\label{eq:phase-def}
\eeq
 The COE (coexistence) phase is favored by the axial anomaly, since the simultaneous
 presence of  $d$ and positive $\sigma$  makes the GL potential
 lower because of the  $\gamma$-term with $\gamma >0$.
 Note that even the unbroken discrete symmetry Z(2) is common between
  CFL and COE phases, so that they cannot be distinguishable
  from the symmetry  point of view.

  For flavor-SU(2) chiral limit with $m_{\rm u,d}=0$ and 
  $m_{\rm s}=\infty$, the condensates with the s quark
  disappear, so that we have 
$  \Phi ={\rm diag}(\sigma,\sigma,0)$ and  
$  \dL=-\dR={\rm diag}(0,0,d)$.
Then 
 the reduced GL potential becomes
$\Omega_{\rm 2F} =
   \left( \frac{a}{2} \sigma^2 + \frac{b}{4} \sigma^4
  + \frac{f}{6} \sigma^6 \right)
  +\left(  \frac{\alpha}{2} {d}^2 + \frac{\beta}{4} {d}^4 \right)
  + {\lambda} {d}^2 \sigma^2$.
 Thus, the coexistence of $d$ and $\sigma$ is disfavored in this case,
  because of the 
  $\lambda$-term with $\lambda >0$.

\subsection{Possible phase structure for realistic quark masses}

\begin{figure}[t]
\begin{center}
\includegraphics[width=0.55\columnwidth]{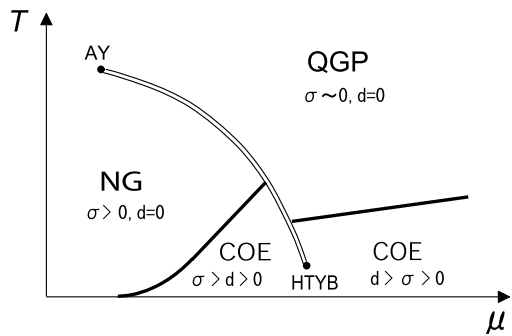}
\end{center}
\caption{Schematic phase structure with two light (up and down) quarks and
a medium heavy (strange) quark \cite{Hatsuda:2006ps}. The double line indicates
 the first order transition.  AY and HTYB are the second-order critical points at which 
  the first-order line terminates.}
\label{fig:32F}
\end{figure}

    The mapping of the phase diagrams obtained from the GL potentials, $\Omega_{\rm 3F}$
     and $\Omega_{\rm 2F}$,
           in the $a-\alpha$ plane to
the $T-\mu$ plane is a dynamical question which cannot be addressed within the
phenomenological GL theory.  Nevertheless,
we can draw a {\em speculative} phase structure of QCD
 for  $m_{\rm s} \sim \Lamqcd \gg m_{\rm u,d} \neq 0$  by interpolating the
 phase structures obtained  from $\Omega_{\rm 3F}$ and $\Omega_{\rm 2F}$
 as shown in Fig.\ref{fig:32F} \cite{Hatsuda:2006ps}. 
    In this figure, the double line indicates the first order phase transition
    driven by the negative $b$ in Eq.(\ref{eq:nf3-model}). The single lines
    indicate the second order phase transitions (within the analysis of the GL potential
    without fluctuations)  which separate the  $d \neq 0$ and $d=0$ phases.
    We draw two critical points at which the first 
    order phase transition turns into crossover;  the  one near the
 vertical axis indicated as ``AY" 
  (Asakawa-Yazaki critical point  \cite{Asakawa:1989bq}) 
 and the other one near the horizontal axis indicated as ``HTYB" \cite{Hatsuda:2006ps}.
 The latter is driven by the axial anomaly with positive $\gamma$  in Eq.(\ref{eq:nf3-model}).  
 
 The existence of the AY critical point implies that the 
  transition from the NG phase to the QGP phase 
 on the $\mu=0$ axis is a crossover.  Indeed,
   the lattice QCD Monte Carlo simulations at finite $T$  
 with the   finite-size scaling analyses  indicate that 
  the thermal phase transition at $\mu=0$ 
  is likely to be crossover \cite{Aoki:2006we}.   
     We note here  that the AY critical point has  special importance to
  the fluctuation observables in relativistic heavy-ion collisions
  \cite{Stephanov:1999zu}
  and the determination of its  location is highly
  called for both theoretically and experimentally. 
 
  On the other hand, the existence of the HTBY critical point
   implies that the haronic matter (characterized by 
   $\sigma > d > 0$  and the quark matter characterized by 
   $d > \sigma > 0$ are    continuously connected with each other
    and both are classified into the COE phase. 
 This  is intimately related to 
  the idea of  hadron-quark continuity, i.e. smooth transition from the 
  superfluid/superconducting hadronic matter  to the superconducting 
 quark matter
  \cite{Schafer:1998ef,Yamamoto:2008zw}.
 Indeed, there are  evidences of the continuity not only for 
 the ground state but also for the excitation spectra: Typical example is  
 the continuity of the  flavor-octet vector mesons in hadronic matter
  at low $\mu$ and  the color-octet gluons in quark matter at high $\mu$ 
   \cite{Schafer:1998ef,Hatsuda:2008is}.
 Unfortunately,       
 the lattice QCD simulations have
  difficulty to treat the matter with $\mu/T \gg 1$ 
  because of the severe sign problem originating from 
  the complex fermion determinant in the presence of $\mu$ \cite{Ejiri:2008nv}.
 Therefore, the quantitative study in this region is still an open issue.

\section{Simulating dense QCD with ultracold atoms}

 Ultracold atomic systems and high density QCD matter, 
 although differing by some twenty orders of magnitude in energy scales, 
 share certain analogous physical aspects, e.g., BEC-BCS crossovers 
 \cite{Baym:2008me}.   Motivated by phenomenological studies 
 of QCD that indicate a strong spin-singlet diquark correlation
  inside the nucleon \cite{Anselmino:1992vg}, we focus here on modeling the 
  transition from the  2-flavor quark matter  
  at high density  to the  nuclear matter at low density in terms of 
  a boson-fermion system, in which small size diquarks are the bosons, 
  unpaired quarks the fermions, and the extended nucleons are regarded 
  as composite boson-fermion particles   \cite{Maeda:2009}.
   This would be a starting point to
  understand the quantum phase transition at $\mu \sim \mu_{\rm c}$
   between the hadronic superfluid 
  discussed in Sec.\ref{sec:SFN} and the color superconductivity discussed
   in Sec.\ref{sec:CSC}.  
   
      Recent advances in atomic 
  physics have made it possible indeed to realize boson-fermion mixture in the
   laboratory.  In particular, tuning the atomic interaction via a 
   Feshbach resonance allows formation of heteronuclear 
molecules, as recently observed in a mixture of $^{87}$Rb and $^{40}$K 
atomic vapors in  a 3D optical lattice \cite{exp1}, and in  an optical 
dipole trap \cite{exp2}. 
 
 Let us start from a non-relativistic boson-fermion mixture with Hamiltonian density,
 \beq
\mathcal{H} \!&=&\! \frac1{2\mb}|\nabla\phi(x)|^2 - \mub
|\phi(x)|^2 + \frac12\bar{g}_{\rm bb}|\phi(x)|^4 \nonumber\\
    \!&+&\! 
       \sum_{\sigma}\biggl(\frac1{2\mf}|\nabla\psi_{\sigma}(x)|^2-\muf
       |\psi_{\sigma}(x)|^2  
       \biggl) 
      + \bar{g}_{\rm ff}|\psi _{\uparrow}(x)|^2  |\psi_{\downarrow}(x)|^2  \nonumber \\
     \!&+&\! \sum_{\sigma}\bar{g}_{\rm bf}|\phi(x)|^2 |\psi_{\sigma}(x)|^2 ,
\label{eq:bf-hamiltonian}
\eeq
where $\phi$ is the boson and $\psi$ the fermion  field. 
 The two internal states of the fermions are labeled by spin indices 
 $\sigma =\{\uparrow ,\downarrow \}$. 
 For simplicity, we consider an equally populated mixture of $n$ bosons and $n$
fermions with  $n_{\uparrow}= n_{\downarrow}=n/2$.  

The bare boson-fermion coupling $\bar{g}_{\rm bf}$ is related to the 
 renormalized coupling $\gbf$ and to
the s-wave scattering length $\abf$ by
\beq
\frac{\mR}{2\pi \abf} 
=\frac{1}{\bar g_{\rm bf} }+
 \int_{|{\bf k}| \le \Lambda}\frac{d^3k}{(2\pi)^3} 
 \frac{1}{\varepsilon_{\mathrm{b}}(k)+ \varepsilon_{\mathrm{f}}(k)},  
\eeq 
where $\varepsilon_{i}(k)= k^2/2m_i$ ($i={\rm b, f}$) is the single-particle kinetic energy,
$\mR$ is the boson-fermion reduced mass, and $\Lambda=\pi/(2 r_0)$ is a high momentum cutoff
 with $r_0$ being a  typical atomic scale.
We assume an attractive bare b-f interaction  ($\bar g_{\rm bf} < 0 $), tunable in magnitude, 
 with $\Lambda $ fixed, so that the scattering length $\abf$ can change sign: 
  $\abf \rightarrow {\bar g_{\rm bf} } \mR/(2\pi)$ for small negative
$\bar g_{\rm bf}$, while $\abf \rightarrow r_0$ for large negative
$\bar g_{\rm bf}$.
We keep the bare boson-boson and fermion-fermion interactions
fixed and repulsive (${\bar g}_{\rm bb} > 0, {\bar g}_{\rm ff} > 0$) 
 for the stability of this system.


 In the regime of the weak bare b-f coupling where the 
 dimensionless parameter $\eta \equiv -1/(n^{1/3}\abf)$  is large and positive,
 the system at low temperature 
 is a weakly interacting  mixture of BEC of the b-bosons  (b-BEC) and 
 degenerate f-fermions. The induced interaction through the 
  density fluctuation of b-BEC  may also lead to the pairing of f's (f-BCS).     
 
 On the other hand, in the regime of strong bare b-f coupling where the 
  $\eta$  is large and negative,
   bound molecules of b-bosons and f-fermions called {\it composite fermions}, N = (bf), are formed 
with a kinetic mass $\mN=\mb+\mf$. 
The s-wave scattering length of two N's of opposite spins can be estimated by the 
 exchange of constituent b or f \cite{Maeda:2009},  
\beq
\aN  \simeq - \frac{\mN}{2\mR} \abf .
\label{eq:anborn}
\eeq
 This is the same in magnitude but is opposite in sign from the scattering length between
difermion molecules due to different  statistics. It can be shown that 
 this result is the leading order term in 
  an extension of the present  model to large internal degrees of freedom.
   
Eq.(\ref{eq:anborn}) implies that the low energy effective interaction
between composite fermions in the spin-singlet channel is weakly attractive; the stronger 
the bare b-f attraction 
the weaker the N-N attraction.  
Such an effective interaction causes composite fermions 
to become BCS-paired (N-BCS) below a transition temperature,
\beq 
T_{\mathrm{c}}(\mathrm{N}\mathchar`- \mathrm{BCS})
&=& \frac{e^\gamma}{\pi}\biggl( \frac{2}{e}\biggl)^{7/3} \varepsilon_{\rm N}
  e^{\pi/(2 k_{\rm F} \aN)} \: .
  \label{eq:TcN}
\eeq
where $\varepsilon_{\rm N}=k_{\rm F} ^2/2\mN$ is the  Fermi energy of the N.

\begin{table}[t]
 \caption{\small Correspondence between the boson-fermion mixture in ultracold atoms
  and the diquark-quark mixture in high density QCD.}
  \begin{center}
  \begin{tabular}{|c|c|}
    \hline 
  cold atoms   & dense QCD \\
   \hline \hline 
b (bosonic atom)  & $d$ (diquark), \\ \hline
f$_{\uparrow, \downarrow}$ (fermionic atom) & $q_{\uparrow, \downarrow}$ (unpaired quark)  
\\ \hline
N$_{\uparrow, \downarrow}$ (boson-fermion molecule) &  ${\cal N}_{\uparrow, \downarrow}$ 
(nucleon)  
\\ \hline
b-f attraction & gluonic attraction  \\ \hline
b-BEC & 2-flavor color superconductivity  \\ \hline
N-BCS & nucleon superfluidity  \\ \hline
  \end{tabular}
 \end{center}
\label{tab:CFA-QCD}
\end{table}

\begin{figure}[t]
\begin{center}
\includegraphics[width=1.0\columnwidth]{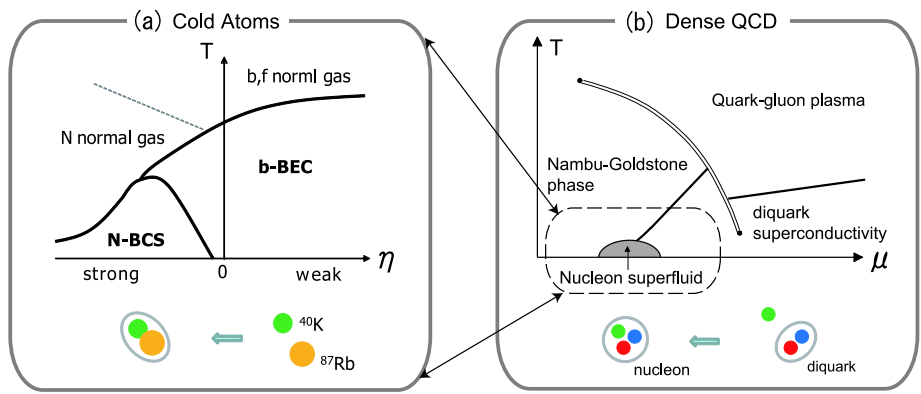}
\end{center}
\caption{(a) A possible phase structure of the boson-fermion mixture (such as 
 $ ^{87}$Rb and $^{40}$K) in ultracold atoms
with attractive b-f interaction and repulsive b-b and f-f interactions.
  Large and positive (large and negative) 
$\eta$  corresponds to the weak (strong) b-f attraction. (b) A  possible
 phase structure of QCD.  Large (small) chemical potential $\mu$ corresponds
  to the weak (strong) coupling due to asymptotic freedom. 
}
\label{Fig4}
\end{figure}

 The above analyses for large  $|\eta|$ 
 suggest a possible phase structures of boson-fermion mixtures 
in the $T-\eta$ plane as shown in Fig.~\ref{Fig4}(a). 
   At intermediate bare b-f coupling ($\eta \sim 0$) where a transition 
from the b-BEC phase to N-BCS takes place, 
the phase diagram would have complex structure
depending on the relative magnitudes of $\bar{g}_{\rm bb}$, $\bar{g}_{\rm ff}$, and 
$\bar{g}_{\rm bf}$.  
The f-BCS phase possibly occurs for $\eta > 0$ is not shown 
in this figure.
 For more detailed analyses of the phase diagram of the present model,
 see \cite{Maeda:2009}.

The phase boundary in the region
 $\eta \sim 0$ may be classified
by the realization of internal symmetry.
If we  focus only on the continuous symmetries,
 the Hamiltonian density, Eq.~(\ref{eq:bf-hamiltonian}), 
has ${\rm U(1)}_{\rm b} \otimes {\rm U(1)}_{\rm f _{\uparrow}} 
\otimes {\rm U(1)}_{\rm f _{\downarrow}}$ 
symmetry corresponding to independent  phase rotations of 
$\phi$, $\psi_{\uparrow}$ and $\psi_{\downarrow}$. 
On the other hand, b-BEC and N-BCS break 
${\rm U(1)}_{\rm b}$
and $U(1)_{\rm b+(f _{\uparrow}+f _{\downarrow})}$ symmetries, respectively. 
The difference in such symmetry breaking patterns
implies the existence of a well-defined phase boundary 
between b-BEC and N-BCS
 as indicated in Fig.~\ref{Fig4}(a).  This is 
  in contrast to the continuous BEC-BCS crossover
   in  two-component Fermi systems.  

The phase structure we find for boson-fermion mixture of ultracold atoms displays
 features of that in  QCD with equal numbers of u and d quarks.
The ground state of such system at high density is the  2-flavor color superconductivity (2SC) 
discussed in Sec.\ref{subsec:gap-eq}.
  The order parameter for color-symmetry breaking
 is the diquark condensate $ \langle {d}_3 \rangle$
 with the diquark operator
  ${d}_{c}=$\\
  $\epsilon_{ij} 
\epsilon_{abc} [q]_{a}^i C \gamma_5 [q]_{b}^j$.
 The  gap is of order a few tens of MeV;
remaining quarks are unpaired 
and form degenerate Fermi seas.  On the other hand, 
the ground state of QCD at low density
is the nuclear matter with equal numbers of protons 
and neutrons denoted by ${\cal N}^i_{\uparrow, \downarrow}$, a superfluid state
with a pairing gap of a few MeV \cite{Dean:2002zx}; 
the order parameter for the spontaneous breaking of 
 baryon-number symmetry ${\rm U(1)}_{\rm B}$ is the 
{\emph six}-quark condensate 
$ \langle {\cal N}^i_{\uparrow} {\cal N}^j_{\downarrow} \rangle 
= \langle ({d}_{a}^{ }[q_{\uparrow}]_a^i)({d}_{b}^{ }[q_{\downarrow}]_b^j) \rangle $.
If we model the nucleon, of radius $r_{_{\cal N}} \sim 0.86 $ fm,
 as a bound molecule of a diquark (of radius $r_{_d} \sim 0.5$ fm) and 
 an unpaired quark, we can make the correspondence between
  boson-fermion mixture of cold atoms and the diquark-quark mixture in 
  QCD as shown in Table \ref{tab:CFA-QCD}.

  Such correspondence can be also found between the phase diagram of
 ultracold atoms in Fig.\ref{Fig4}(a) and that of dense QCD in Fig.\ref{Fig4}(b).
In particular,  
 the BCS-like superfluidity of composite fermion (N) with a small gap is a
 natural consequence of the strong b-f attraction as shown in Eq.(\ref{eq:anborn}), 
 which  may explain why the fermion gap in nucleon superfluidity 
 is order of magnitude smaller than the gap in BEC-like color superconductivity.
 It is thus quite interesting to carry out the experiments of
  boson-fermion mixture in ultracold atoms for wide range of the 
  boson-fermion  attraction.

 Note, however, that  tuning the coupling strength at fixed density
  is not possible in  dense QCD matter because of the 
   running coupling $\als(\kappa=\mu)$; furthermore,   
 dynamical breaking of chiral symmetry and its interplay with the 
 color superconductivity  have an important role in the quantum phase
  transition in QCD as discussed in Sec.\ref{subsec:GL}.
 With these reservations in mind, we suggest that fuller understanding,
  both theoretical and experimental, of the boson-fermion mixture 
  \cite{Maeda:2009}
 as well as a mixture of three species of atomic fermions \cite{Cherng:2007,Rapp:2006rx}
  can  reveal properties of high density QCD
    not readily observable in laboratory experiments.

\section{Conclusion}

 In this Chapter, we have discussed thermal and quantum phase transitions in 
 QCD.  The former is relevant to the physics of hot matter in early universe right 
  after the big bang, while the latter is
  relevant to the physics of dense matter in the interiors of
  neutron stars and quark stars. There are three fundamental QCD phases in the 
   $T - \mu$ plane:
  the NG phase, the QGP phase and the CSC phase (Fig.\ref{fig:T-mu}).
 
 We have shown an interesting possibility of 
  hadron-quark continuity in which the 
 superfluid hadronic phase  and the CSC phase are continuously connected with each other
  at low temperature due to the 
  QCD axial anomaly (Fig.\ref{fig:32F}).
  We have also discussed  that the existence of the 
   nucleon superfluidity 
   may be a logical consequence of the 
   tightly bound diquarks interacting with the unpaired quarks in the CSC phase.
 Although such a system  with $\mu/T \gg 1$
 is difficult to be treated in  
  lattice QCD Monte Carlo simulations at present, a mixture of ultracold atoms with 
  different  masses, different statistics and different internal degrees of freedom
 would provide us with  an exciting new tool  to study the essential features
  of quantum phase transitions in dense QCD (Fig.\ref{Fig4}).

\section*{Acknowledgments}

The authors thank H. Abuki, G. Baym, K. Iida, Y. Nishida, M. Tachibana and  N. Yamamoto
 for useful discussions and collaborations on various subjects covered in this article.
This research  was supported  in part by the
Grant-in-Aid for Scientific Research on Innovative Areas (No. 2004: 20105003) 
and by Japan Society for the Promotion of Science for Young Scientists. 

\newpage
 
\addcontentsline{toc}{chapter}{References}

\end{document}